\documentstyle[twocolumn]{article}
\def\PsfigVersion{1.9}
\ifx\undefined\psfig\else \fi

\let\LaTeXAtSign=\@
\let\@=\relax
\edef\psfigRestoreAt{\catcode`\@=\number\catcode`@\relax}
\catcode`\@=11\relax
\newwrite\@unused
\def\ps@typeout#1{{\let\protect\string\immediate\write\@unused{#1}}}
\ps@typeout{psfig/tex \PsfigVersion}

\def\figurepath{./}

\def\@nnil{\@nil}
\def\@empty{}
\def\@psdonoop#1\@@#2#3{}
\def\@psdo#1:=#2\do#3{\edef\@psdotmp{#2}\ifx\@psdotmp\@empty \else
    \expandafter\@psdoloop#2,\@nil,\@nil\@@#1{#3}\fi}
\def\@psdoloop#1,#2,#3\@@#4#5{\def#4{#1}\ifx #4\@nnil \else
       #5\def#4{#2}\ifx #4\@nnil \else#5\@ipsdoloop #3\@@#4{#5}\fi\fi}
\def\@ipsdoloop#1,#2\@@#3#4{\def#3{#1}\ifx #3\@nnil 
       \let\@nextwhile=\@psdonoop \else
      #4\relax\let\@nextwhile=\@ipsdoloop\fi\@nextwhile#2\@@#3{#4}}
\def\@tpsdo#1:=#2\do#3{\xdef\@psdotmp{#2}\ifx\@psdotmp\@empty \else
    \@tpsdoloop#2\@nil\@nil\@@#1{#3}\fi}
\def\@tpsdoloop#1#2\@@#3#4{\def#3{#1}\ifx #3\@nnil 
       \let\@nextwhile=\@psdonoop \else
      #4\relax\let\@nextwhile=\@tpsdoloop\fi\@nextwhile#2\@@#3{#4}}
\ifx\undefined\fbox
\newdimen\fboxrule
\newdimen\fboxsep
\newdimen\ps@tempdima
\newbox\ps@tempboxa
\long\def\fbox#1{\leavevmode\setbox\ps@tempboxa\hbox{#1}\ps@tempdima\fboxrule
    \advance\ps@tempdima \fboxsep \advance\ps@tempdima \dp\ps@tempboxa
   \hbox{\lower \ps@tempdima\hbox
  {\vbox{\hrule height \fboxrule
          \hbox{\vrule width \fboxrule \hskip\fboxsep
          \vbox{\vskip\fboxsep \box\ps@tempboxa\vskip\fboxsep}\hskip 
                 \fboxsep\vrule width \fboxrule}
                 \hrule height \fboxrule}}}}
\fi
\newread\ps@stream
\newif\ifnot@eof       
\newif\if@noisy        
\newif\if@atend        
\newif\if@psfile       
{\catcode`\%=12\global\gdef\epsf@start{
\def\epsf@PS{PS}
\def\epsf@getbb#1{%
\openin\ps@stream=#1
\ifeof\ps@stream\ps@typeout{Error, File #1 not found}\else
   {\not@eoftrue \chardef\other=12
    \def\do##1{\catcode`##1=\other}\dospecials \catcode`\ =10
    \loop
       \if@psfile
	  \read\ps@stream to \epsf@fileline
       \else{
	  \obeyspaces
          \read\ps@stream to \epsf@tmp\global\let\epsf@fileline\epsf@tmp}
       \fi
       \ifeof\ps@stream\not@eoffalse\else
       \if@psfile\else
       \expandafter\epsf@test\epsf@fileline:. \\%
       \fi
          \expandafter\epsf@aux\epsf@fileline:. \\%
       \fi
   \ifnot@eof\repeat
   }\closein\ps@stream\fi}%
\long\def\epsf@test#1#2#3:#4\\{\def\epsf@testit{#1#2}
			\ifx\epsf@testit\epsf@start\else
\ps@typeout{Warning! File does not start with `\epsf@start'.  It may not be a PostScript file.}
			\fi
			\@psfiletrue} 
{\catcode`\%=12\global\let\epsf@percent=
\long\def\epsf@aux#1#2:#3\\{\ifx#1\epsf@percent
   \def\epsf@testit{#2}\ifx\epsf@testit\epsf@bblit
	\@atendfalse
        \epsf@atend #3 . \\%
	\if@atend	
	   \if@verbose{
		\ps@typeout{psfig: found `(atend)'; continuing search}
	   }\fi
        \else
        \epsf@grab #3 . . . \\%
        \not@eoffalse
        \global\no@bbfalse
        \fi
   \fi\fi}%
\def\epsf@grab #1 #2 #3 #4 #5\\{%
   \global\def\epsf@llx{#1}\ifx\epsf@llx\empty
      \epsf@grab #2 #3 #4 #5 .\\\else
   \global\def\epsf@lly{#2}%
   \global\def\epsf@urx{#3}\global\def\epsf@ury{#4}\fi}%
\def\epsf@atendlit{(atend)} 
\def\epsf@atend #1 #2 #3\\{%
   \def\epsf@tmp{#1}\ifx\epsf@tmp\empty
      \epsf@atend #2 #3 .\\\else
   \ifx\epsf@tmp\epsf@atendlit\@atendtrue\fi\fi}

\chardef\psletter = 11 
\chardef\other = 12

\newif \ifdebug 
\newif\ifc@mpute 
\c@mputetrue 

\let\then = \relax
\def\r@dian{pt }
\let\r@dians = \r@dian
\let\dimensionless@nit = \r@dian
\let\dimensionless@nits = \dimensionless@nit
\def\internal@nit{sp }
\let\internal@nits = \internal@nit
\newif\ifstillc@nverging
\def \Mess@ge #1{\ifdebug \then \message {#1} \fi}

{ 
	\catcode `\@ = \psletter
	\gdef \nodimen {\expandafter \n@dimen \the \dimen}
	\gdef \term #1 #2 #3%
	       {\edef \t@ {\the #1}
		\edef \t@@ {\expandafter \n@dimen \the #2\r@dian}%
		\t@rm {\t@} {\t@@} {#3}%
	       }
	\gdef \t@rm #1 #2 #3%
	       {{%
		\count 0 = 0
		\dimen 0 = 1 \dimensionless@nit
		\dimen 2 = #2\relax
		\Mess@ge {Calculating term #1 of \nodimen 2}%
		\loop
		\ifnum	\count 0 < #1
		\then	\advance \count 0 by 1
			\Mess@ge {Iteration \the \count 0 \space}%
			\Multiply \dimen 0 by {\dimen 2}%
			\Mess@ge {After multiplication, term = \nodimen 0}%
			\Divide \dimen 0 by {\count 0}%
			\Mess@ge {After division, term = \nodimen 0}%
		\repeat
		\Mess@ge {Final value for term #1 of 
				\nodimen 2 \space is \nodimen 0}%
		\xdef \Term {#3 = \nodimen 0 \r@dians}%
		\aftergroup \Term
	       }}
	\catcode `\p = \other
	\catcode `\t = \other
	\gdef \n@dimen #1pt{#1} 
}

\def \Divide #1by #2{\divide #1 by #2} 

\def \Multiply #1by #2
       {{
	\count 0 = #1\relax
	\count 2 = #2\relax
	\count 4 = 65536
	\Mess@ge {Before scaling, count 0 = \the \count 0 \space and
			count 2 = \the \count 2}%
	\ifnum	\count 0 > 32767 
	\then	\divide \count 0 by 4
		\divide \count 4 by 4
	\else	\ifnum	\count 0 < -32767
		\then	\divide \count 0 by 4
			\divide \count 4 by 4
		\else
		\fi
	\fi
	\ifnum	\count 2 > 32767 
	\then	\divide \count 2 by 4
		\divide \count 4 by 4
	\else	\ifnum	\count 2 < -32767
		\then	\divide \count 2 by 4
			\divide \count 4 by 4
		\else
		\fi
	\fi
	\multiply \count 0 by \count 2
	\divide \count 0 by \count 4
	\xdef \product {#1 = \the \count 0 \internal@nits}%
	\aftergroup \product
       }}

\def\r@duce{\ifdim\dimen0 > 90\r@dian \then   
		\multiply\dimen0 by -1
		\advance\dimen0 by 180\r@dian
		\r@duce
	    \else \ifdim\dimen0 < -90\r@dian \then  
		\advance\dimen0 by 360\r@dian
		\r@duce
		\fi
	    \fi}

\def\Sine#1%
       {{%
	\dimen 0 = #1 \r@dian
	\r@duce
	\ifdim\dimen0 = -90\r@dian \then
	   \dimen4 = -1\r@dian
	   \c@mputefalse
	\fi
	\ifdim\dimen0 = 90\r@dian \then
	   \dimen4 = 1\r@dian
	   \c@mputefalse
	\fi
	\ifdim\dimen0 = 0\r@dian \then
	   \dimen4 = 0\r@dian
	   \c@mputefalse
	\fi
	\ifc@mpute \then
		\divide\dimen0 by 180
		\dimen0=3.141592654\dimen0
		\dimen 2 = 3.1415926535897963\r@dian 
		\divide\dimen 2 by 2 
		\Mess@ge {Sin: calculating Sin of \nodimen 0}%
		\count 0 = 1 
		\dimen 2 = 1 \r@dian 
		\dimen 4 = 0 \r@dian 
		\loop
			\ifnum	\dimen 2 = 0 
			\then	\stillc@nvergingfalse 
			\else	\stillc@nvergingtrue
			\fi
			\ifstillc@nverging 
			\then	\term {\count 0} {\dimen 0} {\dimen 2}%
				\advance \count 0 by 2
				\count 2 = \count 0
				\divide \count 2 by 2
				\ifodd	\count 2 
				\then	\advance \dimen 4 by \dimen 2
				\else	\advance \dimen 4 by -\dimen 2
				\fi
		\repeat
	\fi		
			\xdef \sine {\nodimen 4}%
       }}

\def\Cosine#1{\ifx\sine\UnDefined\edef\Savesine{\relax}\else
		             \edef\Savesine{\sine}\fi
	{\dimen0=#1\r@dian\advance\dimen0 by 90\r@dian
	 \Sine{\nodimen 0}
	 \xdef\cosine{\sine}
	 \xdef\sine{\Savesine}}}	      

\def\psdraft{
	\def\@psdraft{0}
}
\def\psfull{
	\def\@psdraft{100}
}

\psfull

\newif\if@scalefirst
\def\psscalefirst{\@scalefirsttrue}
\def\psrotatefirst{\@scalefirstfalse}
\psrotatefirst

\newif\if@draftbox
\def\psnodraftbox{
	\@draftboxfalse
}
\def\psdraftbox{
	\@draftboxtrue
}
\@draftboxtrue

\newif\if@prologfile
\newif\if@postlogfile
\def\pssilent{
	\@noisyfalse
}
\def\psnoisy{
	\@noisytrue
}
\psnoisy
\newif\if@bbllx
\newif\if@bblly
\newif\if@bburx
\newif\if@bbury
\newif\if@height
\newif\if@width
\newif\if@rheight
\newif\if@rwidth
\newif\if@angle
\newif\if@clip
\newif\if@verbose
\def\@p@@sclip#1{\@cliptrue}

\newif\if@decmpr

\def\@p@@sfigure#1{\def\@p@sfile{null}\def\@p@sbbfile{null}
	        \openin1=#1.bb
		\ifeof1\closein1
	        	\openin1=\figurepath#1.bb
			\ifeof1\closein1
			        \openin1=#1
				\ifeof1\closein1%
				       \openin1=\figurepath#1
					\ifeof1
					   \ps@typeout{Error, File #1 not found}
						\if@bbllx\if@bblly
				   		\if@bburx\if@bbury
			      				\def\@p@sfile{#1}%
			      				\def\@p@sbbfile{#1}%
							\@decmprfalse
				  	   	\fi\fi\fi\fi
					\else\closein1
				    		\def\@p@sfile{\figurepath#1}%
				    		\def\@p@sbbfile{\figurepath#1}%
						\@decmprfalse
	                       		\fi%
			 	\else\closein1%
					\def\@p@sfile{#1}
					\def\@p@sbbfile{#1}
					\@decmprfalse
			 	\fi
			\else
				\def\@p@sfile{\figurepath#1}
				\def\@p@sbbfile{\figurepath#1.bb}
				\@decmprtrue
			\fi
		\else
			\def\@p@sfile{#1}
			\def\@p@sbbfile{#1.bb}
			\@decmprtrue
		\fi}

\def\@p@@sfile#1{\@p@@sfigure{#1}}

\def\@p@@sbbllx#1{
		\@bbllxtrue
		\dimen100=#1
		\edef\@p@sbbllx{\number\dimen100}
}
\def\@p@@sbblly#1{
		\@bbllytrue
		\dimen100=#1
		\edef\@p@sbblly{\number\dimen100}
}
\def\@p@@sbburx#1{
		\@bburxtrue
		\dimen100=#1
		\edef\@p@sbburx{\number\dimen100}
}
\def\@p@@sbbury#1{
		\@bburytrue
		\dimen100=#1
		\edef\@p@sbbury{\number\dimen100}
}
\def\@p@@sheight#1{
		\@heighttrue
		\dimen100=#1
   		\edef\@p@sheight{\number\dimen100}
}
\def\@p@@swidth#1{
		\@widthtrue
		\dimen100=#1
		\edef\@p@swidth{\number\dimen100}
}
\def\@p@@srheight#1{
		\@rheighttrue
		\dimen100=#1
		\edef\@p@srheight{\number\dimen100}
}
\def\@p@@srwidth#1{
		\@rwidthtrue
		\dimen100=#1
		\edef\@p@srwidth{\number\dimen100}
}
\def\@p@@sangle#1{
		\@angletrue
		\edef\@p@sangle{#1} 
}
\def\@p@@ssilent#1{ 
		\@verbosefalse
}
\def\@p@@sprolog#1{\@prologfiletrue\def\@prologfileval{#1}}
\def\@p@@spostlog#1{\@postlogfiletrue\def\@postlogfileval{#1}}
\def\@cs@name#1{\csname #1\endcsname}
\def\@setparms#1=#2,{\@cs@name{@p@@s#1}{#2}}
\def\ps@init@parms{
		\@bbllxfalse \@bbllyfalse
		\@bburxfalse \@bburyfalse
		\@heightfalse \@widthfalse
		\@rheightfalse \@rwidthfalse
		\def\@p@sbbllx{}\def\@p@sbblly{}
		\def\@p@sbburx{}\def\@p@sbbury{}
		\def\@p@sheight{}\def\@p@swidth{}
		\def\@p@srheight{}\def\@p@srwidth{}
		\def\@p@sangle{0}
		\def\@p@sfile{} \def\@p@sbbfile{}
		\def\@p@scost{10}
		\def\@sc{}
		\@prologfilefalse
		\@postlogfilefalse
		\@clipfalse
		\if@noisy
			\@verbosetrue
		\else
			\@verbosefalse
		\fi
}
\def\parse@ps@parms#1{
	 	\@psdo\@psfiga:=#1\do
		   {\expandafter\@setparms\@psfiga,}}
\newif\ifno@bb
\def\bb@missing{
	\if@verbose{
		\ps@typeout{psfig: searching \@p@sbbfile \space  for bounding box}
	}\fi
	\no@bbtrue
	\epsf@getbb{\@p@sbbfile}
        \ifno@bb \else \bb@cull\epsf@llx\epsf@lly\epsf@urx\epsf@ury\fi
}	
\def\bb@cull#1#2#3#4{
	\dimen100=#1 bp\edef\@p@sbbllx{\number\dimen100}
	\dimen100=#2 bp\edef\@p@sbblly{\number\dimen100}
	\dimen100=#3 bp\edef\@p@sbburx{\number\dimen100}
	\dimen100=#4 bp\edef\@p@sbbury{\number\dimen100}
	\no@bbfalse
}
\newdimen\p@intvaluex
\newdimen\p@intvaluey
\def\rotate@#1#2{{\dimen0=#1 sp\dimen1=#2 sp
		  \global\p@intvaluex=\cosine\dimen0
		  \dimen3=\sine\dimen1
		  \global\advance\p@intvaluex by -\dimen3
		  \global\p@intvaluey=\sine\dimen0
		  \dimen3=\cosine\dimen1
		  \global\advance\p@intvaluey by \dimen3
		  }}
\def\compute@bb{
		\no@bbfalse
		\if@bbllx \else \no@bbtrue \fi
		\if@bblly \else \no@bbtrue \fi
		\if@bburx \else \no@bbtrue \fi
		\if@bbury \else \no@bbtrue \fi
		\ifno@bb \bb@missing \fi
		\ifno@bb \ps@typeout{FATAL ERROR: no bb supplied or found}
			\no-bb-error
		\fi
		\count203=\@p@sbburx
		\count204=\@p@sbbury
		\advance\count203 by -\@p@sbbllx
		\advance\count204 by -\@p@sbblly
		\edef\ps@bbw{\number\count203}
		\edef\ps@bbh{\number\count204}
		\if@angle 
			\Sine{\@p@sangle}\Cosine{\@p@sangle}
	        	{\dimen100=\maxdimen\xdef\r@p@sbbllx{\number\dimen100}
					    \xdef\r@p@sbblly{\number\dimen100}
			                    \xdef\r@p@sbburx{-\number\dimen100}
					    \xdef\r@p@sbbury{-\number\dimen100}}
                        \def\minmaxtest{
			   \ifnum\number\p@intvaluex<\r@p@sbbllx
			      \xdef\r@p@sbbllx{\number\p@intvaluex}\fi
			   \ifnum\number\p@intvaluex>\r@p@sbburx
			      \xdef\r@p@sbburx{\number\p@intvaluex}\fi
			   \ifnum\number\p@intvaluey<\r@p@sbblly
			      \xdef\r@p@sbblly{\number\p@intvaluey}\fi
			   \ifnum\number\p@intvaluey>\r@p@sbbury
			      \xdef\r@p@sbbury{\number\p@intvaluey}\fi
			   }
			\rotate@{\@p@sbbllx}{\@p@sbblly}
			\minmaxtest
			\rotate@{\@p@sbbllx}{\@p@sbbury}
			\minmaxtest
			\rotate@{\@p@sbburx}{\@p@sbblly}
			\minmaxtest
			\rotate@{\@p@sbburx}{\@p@sbbury}
			\minmaxtest
			\edef\@p@sbbllx{\r@p@sbbllx}\edef\@p@sbblly{\r@p@sbblly}
			\edef\@p@sbburx{\r@p@sbburx}\edef\@p@sbbury{\r@p@sbbury}
		\fi
		\count203=\@p@sbburx
		\count204=\@p@sbbury
		\advance\count203 by -\@p@sbbllx
		\advance\count204 by -\@p@sbblly
		\edef\@bbw{\number\count203}
		\edef\@bbh{\number\count204}
}
\def\in@hundreds#1#2#3{\count240=#2 \count241=#3
		     \count100=\count240	
		     \divide\count100 by \count241
		     \count101=\count100
		     \multiply\count101 by \count241
		     \advance\count240 by -\count101
		     \multiply\count240 by 10
		     \count101=\count240	
		     \divide\count101 by \count241
		     \count102=\count101
		     \multiply\count102 by \count241
		     \advance\count240 by -\count102
		     \multiply\count240 by 10
		     \count102=\count240	
		     \divide\count102 by \count241
		     \count200=#1\count205=0
		     \count201=\count200
			\multiply\count201 by \count100
		 	\advance\count205 by \count201
		     \count201=\count200
			\divide\count201 by 10
			\multiply\count201 by \count101
			\advance\count205 by \count201
		     \count201=\count200
			\divide\count201 by 100
			\multiply\count201 by \count102
			\advance\count205 by \count201
		     \edef\@result{\number\count205}
}
\def\compute@wfromh{
		\in@hundreds{\@p@sheight}{\@bbw}{\@bbh}
		\edef\@p@swidth{\@result}
}
\def\compute@hfromw{
	        \in@hundreds{\@p@swidth}{\@bbh}{\@bbw}
		\edef\@p@sheight{\@result}
}
\def\compute@handw{
		\if@height 
			\if@width
			\else
				\compute@wfromh
			\fi
		\else 
			\if@width
				\compute@hfromw
			\else
				\edef\@p@sheight{\@bbh}
				\edef\@p@swidth{\@bbw}
			\fi
		\fi
}
\def\compute@resv{
		\if@rheight \else \edef\@p@srheight{\@p@sheight} \fi
		\if@rwidth \else \edef\@p@srwidth{\@p@swidth} \fi
}
\def\compute@sizes{
	\compute@bb
	\if@scalefirst\if@angle
	\if@width
	   \in@hundreds{\@p@swidth}{\@bbw}{\ps@bbw}
	   \edef\@p@swidth{\@result}
	\fi
	\if@height
	   \in@hundreds{\@p@sheight}{\@bbh}{\ps@bbh}
	   \edef\@p@sheight{\@result}
	\fi
	\fi\fi
	\compute@handw
	\compute@resv}

\def\psfig#1{\vbox {
	\ps@init@parms
	\parse@ps@parms{#1}
	\compute@sizes
	\ifnum\@p@scost<\@psdraft{
		\special{ps::[begin] 	\@p@swidth \space \@p@sheight \space
				\@p@sbbllx \space \@p@sbblly \space
				\@p@sbburx \space \@p@sbbury \space
				startTexFig \space }
		\if@angle
			\special {ps:: \@p@sangle \space rotate \space} 
		\fi
		\if@clip{
			\if@verbose{
				\ps@typeout{(clip)}
			}\fi
			\special{ps:: doclip \space }
		}\fi
		\if@prologfile
		    \special{ps: plotfile \@prologfileval \space } \fi
		\if@decmpr{
			\if@verbose{
				\ps@typeout{psfig: including \@p@sfile.Z \space }
			}\fi
			\special{ps: plotfile "`zcat \@p@sfile.Z" \space }
		}\else{
			\if@verbose{
				\ps@typeout{psfig: including \@p@sfile \space }
			}\fi
			\special{ps: plotfile \@p@sfile \space }
		}\fi
		\if@postlogfile
		    \special{ps: plotfile \@postlogfileval \space } \fi
		\special{ps::[end] endTexFig \space }
		\vbox to \@p@srheight sp{
			\hbox to \@p@srwidth sp{
				\hss
			}
		\vss
		}
	}\else{
		\if@draftbox{		
			\hbox{\frame{\vbox to \@p@srheight sp{
			\vss
			\hbox to \@p@srwidth sp{ \hss \@p@sfile \hss }
			\vss
			}}}
		}\else{
			\vbox to \@p@srheight sp{
			\vss
			\hbox to \@p@srwidth sp{\hss}
			\vss
			}
		}\fi

	}\fi
}}
\psfigRestoreAt
\let\@=\LaTeXAtSign
 
\begin{document}
{\bf Electron transport in the dye sensitized nanocrystalline
cell}
A Kambili$^a$, A B Walker$^a$, F Qiu$^a$,
A C Fisher$^b$, A D Savin$^b$ and L M Peter$^b$\\
$^a$Department of Physics, $^b$Department of Chemistry,
University of Bath, Bath BA2 7AY, UK

\noindent{\bf Abstract}\\
Dye sensitised nanocrystalline solar cells (Gr\"{a}tzel cells) have achieved
solar-to-electrical energy conversion efficiencies of 12\% in diffuse
daylight. The cell is based on a thin film of dye-sensitised
nanocrystalline TiO$_2$ interpenetrated by a redox electrolyte. The high
surface area of the TiO$_2$ and the spectral characteristics of the dye
allow the device to harvest 46\% of the solar energy flux. 
One of the puzzling features of dye-sensitised nano-crystalline solar
cells is the slow electron transport in the titanium dioxide phase. The
available experimental evidence as well as theoretical considerations
suggest that the driving force for electron collection at the substrate
contact arises primarily from the concentration gradient, ie the
contribution of drift is negligible. The transport of electrons has been
characterised by small amplitude pulse or intensity modulated
illumination.  Here, we show how the transport of electrons in the
Gr\"{a}tzel cell can be described quantitatively using trap
distributions obtained from a novel charge extraction method with a
one-dimensional model based on solving the continuity equation for the
electron density. For the first time in such a model, a back reaction
with the I$_3^-$ ions in the electrolyte that is second order in the
electron density has been included.\\[0.2cm]
\section{Introduction}
\begin{figure}[h]
\centerline{\psfig{figure=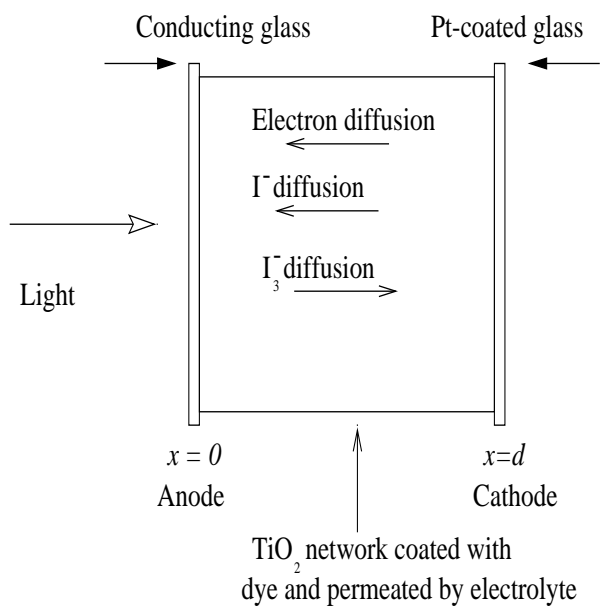,height=8cm,width=8cm}}
\caption{Schematic diagram of solar cell}
\label{fig:fig1}
\end{figure}

Dye-sensitised nanocrystalline cells were developed by Gr\"{a}tzel and
coworkers \cite{oregan,peterreview}. Their operation is shown in
figure \ref{fig:fig1}. Solar radiation is absorbed by a monolayer of a ruthenium
based dye adsorbed on the surface of a porous nanocrystalline film
consisting of electrically connected particles of $\sim$ nm in diameter
of TiO$_2$. On excitation of the dye by the radiation, electrons are
transferred from the HOMO to the LUMO band of the dye, the latter lying
just above the conduction band of the TiO$_2$, and then injected into
the TiO$_2$ on timescales thought to be $\sim$ fs
\cite{rehm,durrant}. The TiO$_2$ film is permeated by an
electrolyte which contains iodide ions (I$^-$) and iodine in the form
I$_3^-$. The dye is regenerated by electron transfer from the I$^-$
ions, leading to the formation of I$_3^-$. In turn, the I$_3^-$ ions are
regenerated by electron transfer from the cathode, facilitated by a
layer of platinum which acts as a catalyst. Electrons injected into the
TiO$_2$ diffuse through the nanocrystalline film to the anode.

Whilst diffusion is slow in the electrolyte, the high concentrations of
I$_3^-$ and I$^-$ ions result in loss free hole transport. Hence much of
the research on these cells has focussed on electron transport in the
nanocrystalline film since injected electrons are slowed down by
trapping at the surface of the TiO$_2$ particles and may back react through
recombination with I$_3^-$ ions. As the surface
area/volume ratio is so high, it seems probable that traps at or near
the surface dominate over those in the bulk. Back reaction with the
dye ions occurs both through geminate reactions and from traps. 
It is likely that back reactions with the dye ions are reduced
by efficient scavenging of the photogenerated holes by the I$^-$ ions
\cite{peterreview}, so back reaction with I$_3^-$
ions will be a significant cause of loss of efficiency.
Measurements of the decay of the total electron concentration at open
circuit show that the rate of back reaction with I$_3^-$ ions is second
order in the total electron concentration
\cite{duffyece,duffyjpc}, and possible back reaction sequences
consistent with this observation were proposed in \cite{fisher}. 

The response of the cell to time varying light intensities can yield
much information about the density of traps, their energy distribution
and the rate and nature of the back reaction. This can be done through
the large amplitude measurements of electron decay in
\cite{duffyece,duffyjpc}, but these have the disadvantage that the whole
trap distribution is sampled in a single measurement. More information
can be obtained by employing intensity modulated photocurrent
spectroscopy (IMPS) and intensity modulated photovoltage spectroscopy
(IMVS) which involve superimposing a small modulated component of light
intensity on a dc component $I_0$ \cite{peterreview,dloczik}. A plot of
the imaginary part of the modulated photocurrent against the real part
of the same quantity gives a semicircle in the negative imaginary half
of the Argand diagram.  The imaginary component is negative as the
response lags behind the light intensity and is the majority carrier
\cite{peterreview}. The magnitude of the imaginary response is a maximum
when the cell is being perturbed at a frequency $\omega_{min}$ which is
the inverse of its response time $\tau$. Here, $\tau$ is the mean
transit time for electrons through the TiO$_2$ network, allowing for
electron scattering which is included in the bare diffusion coefficient
$D_{bare}$ (ie without traps), and multiple trapping.  A similar result
can be derived for the photovoltage.

We have used this approach to find how $\tau$ from IMPS spectra varies
with the background intensity $I_0$. Increasing $I_0$ results in an
increase in the conduction band density and hence the electron
quasi-Fermi level $E_{Fn}$. As  $E_{Fn}$
resides in the bandgap, changing $I_0$ scans $E_{Fn}$ through
the trap levels. Given that the response is dominated by states near
$E_{Fn}$, IMPS and IMVS provide a means of probing the trap density of
states. This was shown \cite{vanmaekelbergh} for a first order
backreaction with the assumption that the gradient of the electron
quasi-Fermi level is constant. 

Below, we show how $\omega_{min}$ can be calculated from a multiple
trapping model in which electron transport is mediated by the conduction
band and is interrupted by trapping. An alternative approach is the
hopping model in which it is assumed that the electrons are always in a
trapped state and transport is achieved by hopping from one trap to
another. Recently, Nelson et al \cite{nelson} looked at both models
using a random walk Monte Carlo model utilising the continuous time
random walk method of Scher and Montroll \cite{sher} where steps to all
nearest neighbours are equally likely and the waiting time depends only
on the energy of the initial site, and which included the back reaction
with I$_3^-$ ions. They showed that the multiple trapping model is more
capable than the hopping model of explaining experimental data on charge
decay due to recombination.

To calculate the IMPS spectra, we solved the continuity equations for
the conduction electron density and trapping probability. Our approach
is a generalisation of that of Vanmaekelbergh and de Jongh
\cite{vanmaekelbergh} where we have included back reactions of first and
second order and avoided the assumption that the gradient of the
electron quasi-Fermi level is constant. 

A common treatment of the transient response, used in
\cite{schlichthorl}, is to include the traps through an effective
diffusion coefficient which depends on the conduction electron density
or equivalently light intensity. For the diffusion coefficient
$D_{bare}$ we have taken the value measured on nanocrystalline anatase
TiO$_2$ which has a small defect concentration. This is because we have
explicitly allowed for the influence of electrolyte and defects in our
model so they should not be also included in the diffusion coefficient.
In \cite{dloczik}, expressions for an effective diffusion coefficient
and effective transition time that explicitly show the influence of the
traps and back reaction were obtained from a similar model to ours but
for a single trap level and a first order back reaction. However, in our
model the physical meaning of these quantities is not clear. 

\section{Method}
\label{sec:method}
The rate of change of electron density $n(x,t)$,
where $x$ is the distance from the anode
and $t$ the time, is determined from
the continuity equation for conduction electrons:
\begin{eqnarray}
\frac{\partial n}{\partial t}= 
D_{bare}\frac{\partial^2 n}{\partial x^2}
+\alpha I_0\exp(-\alpha x)
-k_{cb\mu} (n^{\mu}-n_{dark}^{\mu})\nonumber\\
+<k_{detrap}N_{t0}f-k_{trap}n(1-f)>
\label{eq:cont}
\end{eqnarray} 
Here the first term on the rhs is the photocurrent
due to with  diffusion coefficient without traps $D_{bare}$,
the second term is the generation
rate from light of intensity $I_0$ with an absorption coefficient
$\alpha$, the third term is the back reaction rate for conduction electrons
with the I$_3^-$ ions at a rate of $k_{cb\mu}$, where $\mu$ = 1 for  1st
order recombination, $\mu$ = 2 for 2nd order recombination and
$n_{dark}$ the conduction electron density in the dark, and the fourth
term is the net trapping rate whose terms are defined below. Equation
\ref{eq:cont} is coupled with the continuity equation for trapped
electrons.
\begin{eqnarray}
N_{t0}<\frac{\partial f}{\partial t}>= < k_{trap}n(1-f) 
- k_{detrap}N_{t0}f> \nonumber\\
-k_{tb\mu}(n_{trap}^{\mu}-n_{trapdark}^{\mu})
\label{eq:dfdt}
\end{eqnarray} 
where $N_{t0}$ is the density of traps, $f$ the probabilty that a trap
level is occupied, $k_{trap}$ the trapping rate, $k_{detrap}$ the
detrapping rate and $k_{tb\mu}$ the back reaction rate for trapped
electrons with the I$_3^-$ ions. The angled brackets $<>$ represent 
an average over the trap distribution, for any quantity $A$:
\begin{equation}
<A> \equiv \int_{-E_g}^{0}A(E_T)s(E_T){\rm d}E_T \;.
\end{equation}
where $s(E_T)$d$E_T$ is the probability of a trap site existing
in the energy range $[E_T$, $E_T+$d$E_T]$ in the bandgap $E_g$,
and is normalised so that it integrates to unity over the bandgap.
Thus, the trapped electron density  
\begin{equation}
n_{trap} = N_{t0}< f >\;,
\end{equation}
and $n_{trapdark}$ is $n_{trap}$ in the dark. Detrapping 
by thermal excitation is assumed, so
\begin{equation}
k_{detrap} =\frac{k_{trap}N_c}{N_{t0}}\exp\left\{\frac{E_T}{k_BT}\right\}
\;,
\label{eq:kdetrap}
\end{equation}
where $k_B$ is the Boltzmann constant.

An exponential trap distribution was used here in common with other
authors (\cite{nelson,vandelagemaat}), using the parameter $\beta$ to
represent the width of the distribution in units of the thermal energy
$k_BT$ :
\begin{equation}
s(E_T)=\frac{\beta}{k_BT}\exp\left[\frac{\beta E_T}{k_BT}\right]
\label{eq:strap}
\end{equation}
If $\beta$ is large, say 0.3, most of the traps sit up near the
conduction band, so detrapping will take place on average relatively
quickly. As $\beta$ is reduced, the probability of having deeper trap
levels increases corresponding to longer detrapping times. The limits $\beta
\rightarrow \infty$ and $\beta \rightarrow 0$ correspond respectively to
all traps situated just below the conduction band and to a uniform trap
distribution throughout the bandgap. We have also looked at Gaussian
distributions as recent transient current measurements \cite{wang}
suggest that the trap distribution may be peaked but these results will
be published separately \cite{walker}.

The photocurrent density 
\begin{equation}
j_{photo}(x) = qD_{bare}\frac{\partial n}{\partial x}
\end{equation}
where $q$ is the magnitude of the electron charge.
At the anode,
\begin{equation}
j_{photo}(0) = qk_{ext}(n(0)-n_{dark})\;,
\end{equation}
where $k_{ext} =\sqrt{k_BT/(2m_n^*\pi)}$ for IMPS spectra (short
circuit) and $k_{ext} =0$ for IMVS spectra (open circuit). This is the
Schottky boundary condition \cite{sze} with a recombination velocity
$v_r=k_{ext}$. At the cathode side of the TiO$_2$ network, 
\begin{equation}
j_{photo}(d)= 0 \;.
\end{equation}

The incident photon to electron conversion efficiency IPCE is
$j_{photoss}(x=0)/qI_0$ where $j_{photoss}$ is the steady state
photocurrent density. If trapping effects are neglected, it can be
shown from equation \ref{eq:cont} that the IPCE is equal to
$1-\exp(-\alpha d)$. The ac photocurrent conversion efficiency is
$\Phi(\omega)=j_{photo}(x=0)/q\delta I_0$. We solved equations
\ref{eq:cont} and \ref{eq:dfdt} in the steady state and for the
modulated intensity $\delta I_0 \exp(i\omega t)$  (where $\delta I_0
\sim 0.01 I_0$ so that only the term linear in $\delta I$ was included).
Hence $\Phi(\omega)$ is complex.

As the electrostatic potential is screened out, classical statistics gives
\begin{equation}
n=N_c\exp(E_{Fn}/k_BT)
\end{equation}
where the conduction band density of states
\begin{equation}
N_c = 2 \left[ \frac{m_n^* k_B T}{2\pi\hbar^2} \right]^{3/2}\;
\end{equation}
where $m_n^*$ is the effective mass and $\hbar$ is Planck's constant
divided by 2$\pi$.
 
To solve the equations, a relaxation method for a two point boundary
value problem \cite{press} was adopted. Second order kinetics resulted
in nonlinear functional equations for the steady state and modulated
components of $f$, which were solved using the Newton-Raphson method
\cite{press}. 

The parameter values are given in Table \ref{tab:tab1}.
\begin{table}
\caption{Parameter values}
\begin{tabular}{ccc}
\hline
Symbol & Value\\
\hline
$D_{bare}$ & 1.0 $\times$ 10$^{-6}$ m$^2$s$^{-1}$\\
$m_n^*$ & 9$m_0$\\
$N_c$ & 6.775$\times$ 10$^{26}$ m$^{-3}$\\
$\alpha$ & 2.3 $\times$ 10$^{5}$ m$^{-1}$\\
$T$ & 300 K\\
$E_{Fndark}$ & -1.0 eV\\
$k_{trap}$ & 1.0 $\times$ 10$^7$ s$^{-1}$\\
$d$ & 6 $\times$ 10$^{-6}$m\\
$k_{cb2}=k_{tb2}$ & 2 $\times$ 10$^{-25}$m$^3$s$^{-1}$\\
$\beta$ & 0.2\\
\hline
\end{tabular} 
\label{tab:tab1}
\end{table}
$D_{bare}$ and $m_n^*$ (in units of the bare electron mass $m_0$) were
taken from ref \cite{enright}. We deduced $\alpha$ by assuming an IPCE
of 0.75 \cite{franco}. Our dark Fermi level $E_{Fndark}$
comes from an estimate of 10$^{11}$ m$^{-3}$ for $n_{dark}$ \cite{fisher}.
For an estimate for $k_{trap}$, we assumed that the trap cross-section
$\sigma$ would be in the region of 2$\times$10$^{-22}$m$^{-2}$, and that
$k_{trap}=N_{t0}\sigma v_{th}$ where the thermal velocity of the
electrons $v_{th}=\sqrt{2\pi}v_r$, but to keep the model simple, we made
$k_{trap}$ independent of $N_{t0}$. Our chosen values for
$k_{cb2}$,$k_{tb2}$ and $\beta$ are discussed  at the beginning of
section \ref{sec:results}.
\section{Results}
\label{sec:results}
Our results are shown for a limited range of values of the trap density
$N_{t0}$, trap distribution width $\beta$ and back reaction rate
$k_{tb2}$. We have set $k_{tb2}$=$k_{cb2}$ on the grounds of simplicity.
Our results are insensitive to $k_{cb2}$ as most
of the electrons are trapped. We have chosen a value for $k_{tb2}$ 100
times less than the value deduced from experiment \cite{duffyjpc} as
this value was found to give better agreement between our IMPS
predictions and experimental results. 

For the parameter values shown, our calculations of $\omega_{min}$ give
the best agreement with experiment for a range of values. How our
predicted spectra vary with $\beta$ and $k_{tb2}$ will be presented in a
separate paper \cite{walker}, where our results for the IPCE as a
function of intensity will also be given. We have not attempted to find
a set of values of $N_{t0}$,$k_{tb2}$ and $\beta$ to give the closest
fit to experimental data as the model does not seem accurate enough to
justify this, and it is more instructive to see the trends in
$\omega_{min}$as $N_{t0}$,$k_{tb2}$ and $\beta$ are varied. We chose the
values of $N_{t0}$ from noting that the density of states obtained from
charge extraction measurements \cite{duffyece} is consistent with an
exponential distribution of the type given in equation \ref{eq:strap}
for which $N_{t0}$ = 2.3$\times$10$^{24}$m$^{-3}$ and $\beta$=0.14
\cite{walkerunpub}. However, more recent data suggests that $\beta$ may
be as high as 0.25 \cite{duffyunpub}. Our chosen value for $\beta$ lies
within this range, and it is at the bottom of the range of values
suggested by Nelson et al \cite{nelson} when modelling the decay of
photoinduced dye cations.
\subsection{Steady State Results}
\begin{figure}[h]
\centerline{\psfig{figure=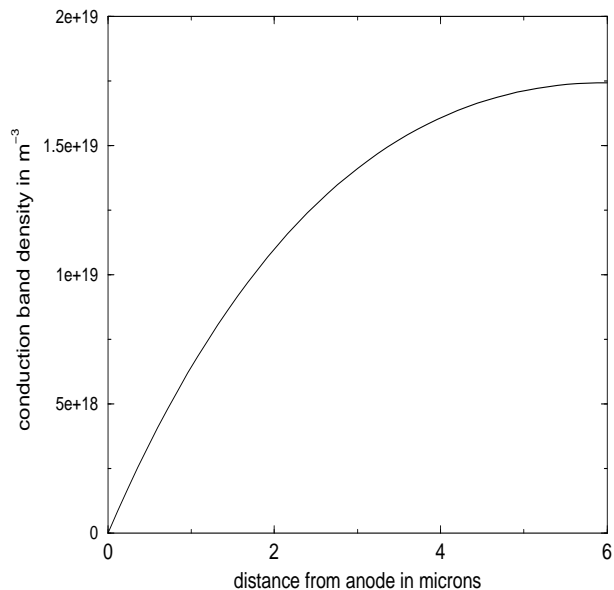,height=8cm,width=8cm}}
\caption{Conduction band density profile $n(x)$ where
$x$ is the distance from the anode for an
intensity $I_0$ of 1$\times$10$^{19}$m$^{-2}$s$^{-1}$ and a trap density
$N_{t0}$ of 2.3$\times$10$^{24}$m$^{-3}$}
\label{fig:fig2}
\end{figure}

These results were obtained by setting the lhs of equations
\ref{eq:cont} and \ref{eq:dfdt} to zero.
In figure \ref{fig:fig2}, we show an example profile for the conduction
band density $n(x)$. This figure demonstrates the large density gradient
that drives the electrons to the anode. The photocurrent is determined
by the trap distribution at the anode where the electron density is much
lower  compared with its value on the cathode side. This means that
there will be significant differences compared to IMVS spectra at the
same intensity, for which $n(x)$ is almost constant  throughout the
TiO$_2$ film because of the boundary conditions forcing the gradient of
$n(x)$ to be zero on both sides of the film.   As the profiles for
$n(x)$ are the same shape at all intensities and $n(x=0)$ is the only
value needed for the photocurrent and IMPS results, we have only given
our results for $n(x=0)$ which are shown in Table  \ref{tab:tab2}.
\begin{table}
\caption{Values $n(x=0)$ at different background intensities 
$I_0$ (m$^{-2}$s$^{-1}$) and trap densities $N_{t0}$(m$^{-3}$)}
\begin{tabular}{ccc}
\hline
$N_{t0}$ & $I_0$=1$\times$10$^{15}$ 
 & $I_0$=1$\times$10$^{19}$\\
   2.3$\times$10$^{23}$ 
 & 9.41$\times$10$^{10}$ 
 & 8.35$\times$10$^{14}$\\
   2.3$\times$10$^{24}$ 
 & 8.53$\times$10$^{10}$ 
 & 8.34$\times$10$^{14}$\\
\hline
\end{tabular} 
\label{tab:tab2}
\end{table}
These results show that $n(x=0)$ is nearly proportional to the light
intensity, so that the IPCE is almost constant as would be expected
for the low back reaction chosen. It is interesting to note that
$n(x=0)$ is insensitive to the trap density at both light intensities. 
 
The difference in behaviour for the different trap densities
is rather seen in the steady state values for
$f$ at the anode which are shown in figure \ref{fig:fig3} for the same trap densities
and light intensities as in Table \ref{tab:tab2}. 
This graph shows the expected Fermi-Dirac distribution at the lower
intensity, although for the lower trap density the spread in $f$ is greater
than for the higher trap density. This leads to a larger $n_{trap}$ and
hence back reaction rate compared to the solution for the higher trap
density. From  equation \ref{eq:dfdt}, setting the left hand side to
zero in the steady state, it can be seen that a large back reaction
forces $f$ to be less than one at low trap energies, and this effect
increases, the more $n_{trap}$ exceeds $n_{trapdark}$. The same
phenomenon is seen at the higher trap density but here the
difference between $n_{trap}$ and $n_{trapdark}$ is much greater at
this density, forcing the observed low value of $f$ at low
trap energies.   

The plateau between trap energies of -0.7 and -0.9 eV at the higher
intensity and trap density resembles closely the plateaux predicted by
Simmons and Taylor \cite{simmons} in calculations of $f$ in the steady
state for bipolar semiconductors with traps. Here, we have the same
system except that it is unipolar and there is a second order back
reaction, so the fact that we see the same features is reassuring.  What
is different from reference \cite{simmons} is our result that $f$ is
less than unity even at trap levels deep into the conduction band. This is a
direct consequence of the assumption that the  back reaction rate is
independent of the trap energy $E_T$, and may therefore be an artifact
of the model. However, it does not unduly affect the results as what is
important is the overlap with the trap distribution $s(E_T)$ and this
decreases rapidly for low values of $E_T$. 
\subsection{IMPS Results}
Figure \ref{fig:fig4} compares the predicted
IMPS spectra for two light intensities. The
spectrum is approximately semicircular. Im($\Phi(\omega_{min})$) at
the higher intensity is about three times its value at the lower
intensity. This can be qualitatively explained as being due to the 
greater losses due to the back reaction at the higher intensity.

\begin{figure}[h]
\centerline{\psfig{figure=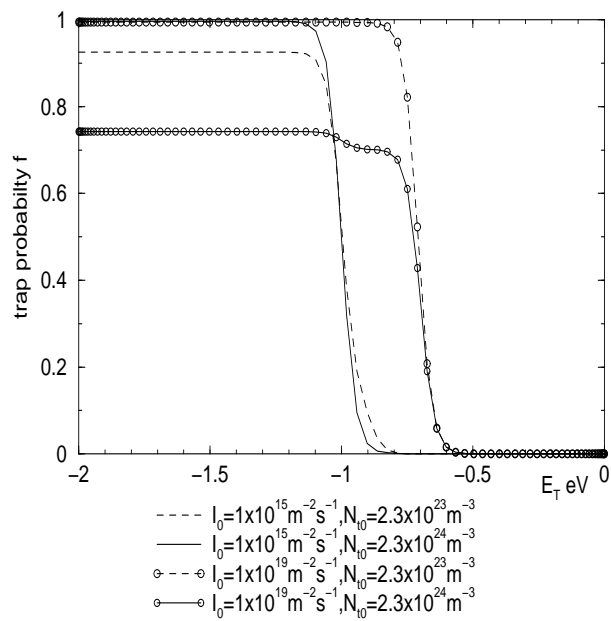,height=8cm,width=8cm}}
\caption{Trap occupation probability $f$ as a function of the trap
energy relative to the conduction band energy $E_T$ for 
intensities $I_0$ of 1$\times$10$^{15}$m$^{-2}$s$^{-1}$ 
and 1$\times$10$^{19}$m$^{-2}$s$^{-1}$ and trap densities
$N_{t0}$ of 2.3$\times$10$^{23}$m$^{-3}$ and
2.3$\times$10$^{24}$m$^{-3}$}
\label{fig:fig3}
\end{figure}

\begin{figure}[h]
\centerline{\psfig{figure=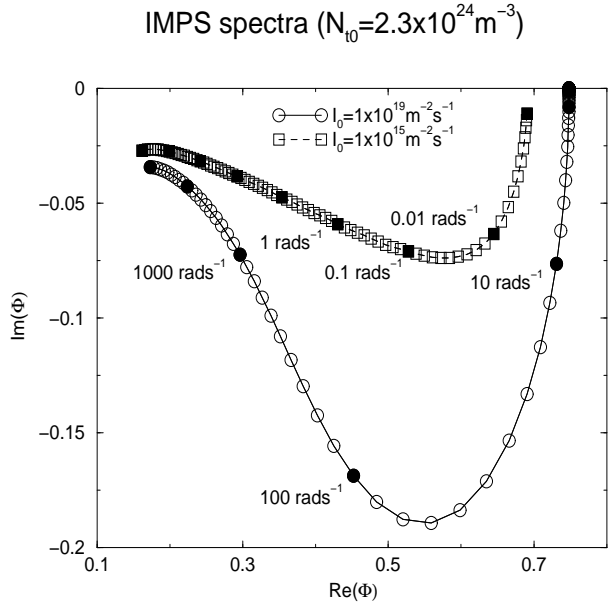,height=8cm,width=8cm}}
\caption{Imaginary part vs real part of the ac photocurrent conversion
efficiency $\Phi$ at intensities $I_0$ of
1$\times$10$^{15}$m$^{-2}$s$^{-1}$ (squares) and
1$\times$10$^{19}$m$^{-2}$s$^{-1}$ and a trap density $N_{t0}$ of
2.3$\times$10$^{24}$m$^{-3}$ (circles)}
\label{fig:fig4}
\end{figure}

The semicircle is a result of a single value for $\tau$ even though
there is a distribution of trap levels; Vanmaekelbergh et al
\cite{vanmaekelbergh} note that this is a result of the capture rate for
a given trap level varying inversely with the residence time. However at
the  higher intensity, where the back reaction is more important,
the semicircle is more distorted as seen from figure \ref{fig:fig3}. This
distortion suggests that there is a dispersion in the electron arrival
times at the anode \cite{vanmaekelbergh}. For each spectrum, it can be
seen that there is a single value of $\omega_{min}$ within the range of
frequencies calculated. Whilst another value exists at much higher
frequencies, this has not been calculated as it would occur well outside
the frequencies accessible to experiment. 

\begin{figure}[h]
\centerline{\psfig{figure=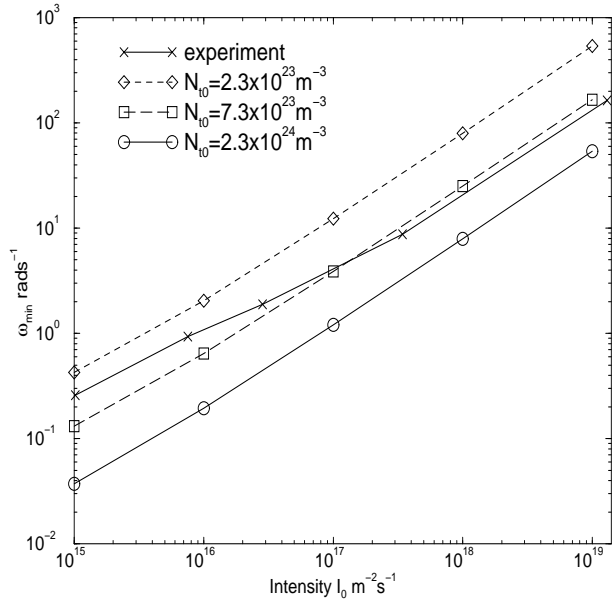,height=8cm,width=8cm}}
\caption{Minimum frequency $\omega_{min}$ vs background intensity
$I_0$ for trap densities $N_{t0}$ of 2.3$\times$10$^{23}$m$^{-3}$
(diamonds), 7.3$\times$10$^{23}$m$^{-3}$ (squares) and
2.3$\times$10$^{24}$m$^{-3}$ (circles) compared with experiment
(crosses)} 
\label{fig:fig5}
\end{figure}

Our main result is shown in figure \ref{fig:fig5} where our predictions
for $\omega_{min}$ are compared with experimental data from
\cite{peterece}. The magnitude of $\omega_{min}=1/\tau$ is determined by
the time taken in trapping and detrapping processes. As $I_0$ increases,
the traps fill up (see figure \ref{fig:fig3}) so that detrapping occurs
from the shallower trap levels at which it takes place faster from
equation \ref{eq:kdetrap}. From figure \ref{fig:fig5}, it is clear that
at any given intensity over the range considered, $\omega_{min}$
decreases as $N_{t0}$ increases. This is because if there are more
traps, the likelihood of an electron being trapped and then detrapped en
route to the anode increases, slowing it down so increasing $\tau$. 

The good agreement with experiment is encouraging. In our calculations
for all three trap densities, we have achieved the same gradient of
approx 0.8 for $\log_{10}(\omega_{min})$ vs $\log_{10}(I_0)$ as in the
experiments at the higher intensities. However, unlike our calculations,
the experimental results flatten off at the lower intensities. This
effect seen in our calculations for larger values of $k_{tb2}$, although
in our calculations the flattening off exceeds that seen experimentally.
The flattening occurs because the trap probability in the steady state
$f$ does not differ greatly from its dark value except at the higher
intensities. 
\section{Conclusions}
\label{sec:conclusions}
We have shown for the first time with a one-dimensional model including
explicitly trapping and back reaction with I$_3^-$ ions that the
assumption of a back reaction which is second order in the electron
density gives values of $\omega_{min}$ deduced from IMPS spectra which
are in good agreement with experiment. Our model will be useful in
obtaining quantitative information from intensity modulated spectra on
the trap distribution and back reaction and hence establish how
different methods of cell fabrication yield different conversion
efficiencies. Many questions however remain unanswered, for example
whether the same parameters will give equally good agreement with IMVS
spectra and whether a suitable set of parameters could give good
agreement with experiment if a first order back reaction were assumed.

\end{document}